\documentclass[lettersize,conference]{template/IEEEtran/IEEEtran}

\usepackage{amsmath,amsfonts,epsfig}
\usepackage{hyperref}
\usepackage{multirow}
\usepackage{graphicx}
\usepackage{float}
\usepackage{fancyhdr}
\usepackage{cite}
\usepackage{xifthen}
\usepackage{makecell}
\usepackage{booktabs}
\usepackage{fancyhdr}
\begin{document}


%
\title{ NN-VVC: Versatile Video Coding boosted by self-supervisedly learned image coding for machines 
}
\newcommand*\samethanks[1][\value{footnote}]{\footnotemark[#1]}
\author{
    \IEEEauthorblockN{
      Jukka I. Ahonen \IEEEauthorrefmark{1}\IEEEauthorrefmark{3}\IEEEauthorrefmark{2},
      Nam Le\IEEEauthorrefmark{1}\IEEEauthorrefmark{3}\IEEEauthorrefmark{2},
      Honglei Zhang\IEEEauthorrefmark{1},
      Antti Hallapuro\IEEEauthorrefmark{1},
      Francesco Cricri\IEEEauthorrefmark{1}, \\
      Hamed Rezazadegan Tavakoli\IEEEauthorrefmark{1},
      Miska M. Hannuksela\IEEEauthorrefmark{1},
      Esa Rahtu\IEEEauthorrefmark{3}
      }\\
    \IEEEauthorblockA{\{jukka.1.ahonen, nam.le, honglei.1.zhang, antti.hallapuro, francesco.cricri, \\hamed.rezazadegan\_tavakoli, miska.hannuksela\}@nokia.com, esa.rahtu@tuni.fi}
    \IEEEauthorblockA{\IEEEauthorrefmark{1}Nokia Technologies, \IEEEauthorrefmark{3}Tampere University, \IEEEauthorrefmark{2}Equally contributed}


}






\maketitle

\newcommand{\newparagraph}[1]{\par\textbf{#1}} 
\renewcommand*{\figureautorefname}{Fig.} 

    \newcommand{\Tensor}[1]{\boldsymbol{#1}} 
    \newcommand{\Loss}[1]{\mathcal{L}_{#1}} 
    \newcommand{\LossFT}[1]{\bar{\mathcal{L}}_{#1}} 
    \newcommand{\Wof}[1]{\bar{w}_{#1}} 
    \newcommand{\Model}[1]{\boldsymbol{#1}} 
    \newcommand{\ModelWeights}[1]{\boldsymbol{\theta}_{\boldsymbol{#1}}} 
    \newcommand{\Wmodel}[2]{\Model{#1}({#2};\ModelWeights{#1})} 
    \newcommand{\goodresult}[1]{\textcolor{green}{#1}} 
    \newcommand{\badresult}[1]{\textcolor{red}{#1}} 
    \newcommand{\expnum}[1]{\mathrm{e}{#1}} 

        
        \newcommand{\litencoder}{E}
        \newcommand{\litdecoder}{D}
        \newcommand{\litprobmodel}{P}
        \newcommand{\litquantizer}{Q}
        \newcommand{\litimg}{x}
        \newcommand{\litlatent}{y}
        \newcommand{\lithyperlatent}{z}
        \newcommand{\litprior}{p}
        \newcommand{\litweight}{w}
        \newcommand{\litweightset}{W}
        \newcommand{\litlossrate}{rate}
        \newcommand{\litlosstask}{task}
        \newcommand{\litlossmse}{mse}
        \newcommand{\litlossfinetune}{total}
        \newcommand{\litlossproxy}{proxy}

        \newcommand{\img}{\Tensor{\litimg}}
        \newcommand{\resimg}{\Tensor{\hat{\litimg}}}
        \newcommand{\latent}{\Tensor{\litlatent}}
        \newcommand{\iqlatent}{\Tensor{\hat{\litlatent}}} 
        \newcommand{\tqlatent}{\Tensor{\Tilde{\litlatent}}} 
        \newcommand{\hyperlatent}{\Tensor{\lithyperlatent}}
        \newcommand{\tqhyperlatent}{\Tensor{\Tilde{\lithyperlatent}}}
        \newcommand{\iqhyperlatent}{\Tensor{\hat{\lithyperlatent}}}
        \newcommand{\prior}[1]{\litprior_{#1}}
        \newcommand{\encoder}[1]{\Wmodel{\litencoder}{#1}}
        \newcommand{\decoder}[1]{\Wmodel{\litdecoder}{#1}}
        \newcommand{\probmodel}[1]{\Wmodel{\litquantizer}{#1}}
        \newcommand{\wrate}{\litweight_{\litlossrate}} 
        \newcommand{\wtask}{\litweight_{\litlosstask}}
        \newcommand{\wmse}{\litweight_{\litlossmse}}
        \newcommand{\wproxyA}{\litweight_{\litlossproxy_A}}
        \newcommand{\wset}{\mathcal{\litweightset}}
        \newcommand{\lossrate}{\Loss{\litlossrate}}
        \newcommand{\losstask}{\Loss{\litlosstask}}
        \newcommand{\lossmse}{\Loss{\litlossmse}}
        \newcommand{\lossproxy}{\Loss{\litlossproxy}}
        \newcommand{\etal}{\textit{et al.}}
    

\begin{abstract}
The recent progress in artificial intelligence has led to an ever-increasing usage of images and videos by machine analysis algorithms, mainly neural networks. Nonetheless, compression, storage and transmission of media have traditionally been designed considering human beings as the viewers of the content. Recent research on image and video coding for machine analysis has progressed mainly in two almost orthogonal directions. The first is represented by end-to-end (E2E) learned codecs which, while offering high performance on image coding, are not yet on par with state-of-the-art conventional video codecs and lack interoperability. The second direction considers using the Versatile Video Coding (VVC) standard or any other conventional video codec (CVC) together with pre- and post-processing operations targeting machine analysis. While the CVC-based methods benefit from interoperability and broad hardware and software support, the machine task performance is often lower than the desired level, particularly in low bitrates. This paper proposes a hybrid codec for machines called NN-VVC, which combines the advantages of an E2E-learned image codec and a CVC to achieve high performance in both image and video coding for machines. Our experiments show that the proposed system achieved up to \(-43.20\%\) and \(-26.8\%\) Bj{\o}ntegaard Delta rate reduction over VVC for image and video data, respectively, when evaluated on multiple different datasets and machine vision tasks. To the best of our knowledge, this is the first research paper showing a hybrid video codec that outperforms VVC on multiple datasets and multiple machine vision tasks. 
\end{abstract}

\section{Introduction}
\IEEEPARstart{I}{mage} and video data consumed by machines have been increasing rapidly in recent years. In this paper, we refer to machines as any algorithm that may analyze an input image or video, in order to obtain analysis results. Examples include object detection, image segmentation, instance segmentation, object tracking, person tracking, etc. Cisco Annual Internet Report \cite{cisco_annualinternet} gave an estimate that by the year 2023, half of the internet traffic will be solely between machines. Thus, it is highly desired to compress images and videos targeted to machine consumption more efficiently than when applying traditional codecs for the benefits in terms of bandwidth savings. 
Thus, the Video Coding for Machines (VCM) Ad-hoc group of Moving Picture Experts Group (MPEG) \cite{cit:vcm}, as well as the JPEG-AI group of JPEG \cite{cit:jpeg-ai} have been actively investigating new technologies for machine-oriented image and video coding standardization. In this regard, the VCM group lists some of the most important use cases in one of their documents \cite{vcm_requirements_N190}, which include surveillance, intelligent transportation, smart cities, intelligent industry, intelligent content, and consumer electronics, which all have a demand for efficient image and video codecs specifically tailored for machines. While existing state-of-the-art codecs such as the High Efficiency Video Coding (HEVC) \cite{HEVC} or the Versatile Video Coding (VVC) \cite{vvc} may be used for machine vision tasks, they are ultimately developed to optimize the compression gains for humans as the end user and are not the most optimal solution when the end user is a machine. 
Responses to the call for evidence (CfE) and to the call for proposals (CfP) of MPEG VCM have shown new technologies that compress images and videos targeted for machine consumption much more efficiently than traditional video codecs, such as VVC, optimized for user viewing. 

In this paper, we present a complete system for compressing images and videos for machines. The proposed system was submitted to the MPEG VCM as a CfP response and it is being studied as a prominent candidate. We combine an end-to-end self-supervisedly learned intra-frame codec with a conventional inter-frame codec, in order to leverage the benefits of these two approaches. Thanks to this combination, we are able to achieve substantial coding gains over VVC in all tested datasets for machine consumption. 

The paper is organized as follows: Section 2 reviews the prior works on learned codecs and in general on video codecs for machines; Section 3 describes the details of the proposed codec; Section 4 provides information on experimental setup and results, including ablation studies; finally, Section 5 draws the conclusions of our paper. 
\label{sec:intro}

\section{Related work}
\label{sec:related-work}
Since the rise of end-to-end (E2E) learned image codecs \cite{duanEndtoEndImageCompression2022, jpegai_paper, liVariableRateDeepImage2022, ho2021anfic,cheng2020learned,  scale_hyperprior,meanscale_hyperprior} that rival or outperform the state-of-the-art traditional codecs HEVC \cite{HEVC} and VVC \cite{cit:vvc} in terms of rate--distortion trade-off according to many quality metrics and coding conditions, video coding with neural network (NN) based components has been an attractive research topic. End-to-end learned video codecs have been explored for the possibility they would inherit the success of the learned image codecs. Agustsson \etal \cite{agustsson2020scale} introduced \textit{scale field} as an additional flow field dimension for more flexibility in motion compensation. Furthermore, in \cite{aivc} and \cite{mentzer2022vct}, motion compensation is handled by conditional autoencoder and Transformer \cite{transformer}-based components, respectively.
In a different aspect, the authors of \cite{choiAffineTransformationBasedDeep2021,clic_paper} seek to enhance the traditional codecs with the aid of NN-based modules, in particular for frame prediction. In comparison, our proposed method instead aims to harmonize the procedures of state-of-the-art conventional video codec and E2E learned image codec to achieve consistent gains in a wide range of test cases and input data.

Similar to the achievements of neural networks in the codecs for human vision, superiority against traditional codecs has been observed in the field of Image coding for Machine vision (ICM). Le \etal \cite{icassp_paper,icme_paper} proposed an E2E learned image codec and domain adaptation techniques that save almost half of the bitstream size compared to the state-of-the-art video codec VVC. The authors of \cite{vcm_feature_based,fischer2022quantizationscaling} proposed rate-distortion optimization methods for the VVC to achieve further coding gains. On the other hand, for many applications, offloading the computational demands on the cloud is crucial. To achieve that, instead of compressing images and videos directly,  the devices on the video-acquiring side may extract intermediate features from the input pictures, compress the features into a bitstream, and send the bitstream to the cloud for further analysis. This compression technique is known as ``feature compression'' for machines. Yamazaki \etal \cite{yamazakiDeepFeatureCompression2022} presented an E2E learned system for feature compression, while \cite{feature_residuals_seppala2021,chenNewImageCodec2021, choiScalableImageCoding2022} proposed scalable image feature coding schemes for human and machine visions. With the advantage of optimizing the codec with the targeting task network, these coding methods achieve significant gains over the VVC codec.
The authors of \cite{choiScalableVideoCoding2022, huangHMFVCHumanMachineFriendly2022} extended the scalable feature coding approach to the video domain. Our work, in contrast to ``feature compression'', operates on the picture domain, i.e., it encodes and decodes pictures and task networks take pictures as their input.

Another related topic to enhance the coding efficiency is NN-based filters. Generally, they can be divided into two categories, in-loop filters  \cite{in-loop_IDAM, in-loop_adaptive_RL, in-loop_contentaware} and post-processing filters \cite{pfilter_ahonen2021, PF_selection, PF_multiheadreso}. An in-loop filter is located inside the codec and processes an input picture to generate an enhanced picture which is often used as a reference for other pictures. A post-processing filter is located after the main decoder and enhances the reconstructed output picture. In order to make the NN filters adaptive to the input content, authors in \cite{LamAdaptivePF} first trained an NN-based post-processing filter. At the inference stage, the pretrained filter is finetuned by overfitting the bias terms of the decoder by minimizing the rate-distortion loss given an input content.  In addition to the finetuning concept, the authors in \cite{mariaAdaptivePF, mariaAdaptivePF_NNRsignal} proposed to finetune scaling factors that determine the strength of the filtering. Authors of \cite{QPAdaptiveInloop1, QPAdaptiveInloop2} proposed using additional information such as quantization parameters for modulating the input features of the NN-layers during adaptation. 

\section{Proposed method}
\label{sec:method}

\subsection{The NN-VVC system}
\label{ssec:NNVVC}
\begin{figure*}[ht]
  \begin{minipage}[b]{\linewidth}
    \centering
      \centerline{\includegraphics[width=\linewidth]{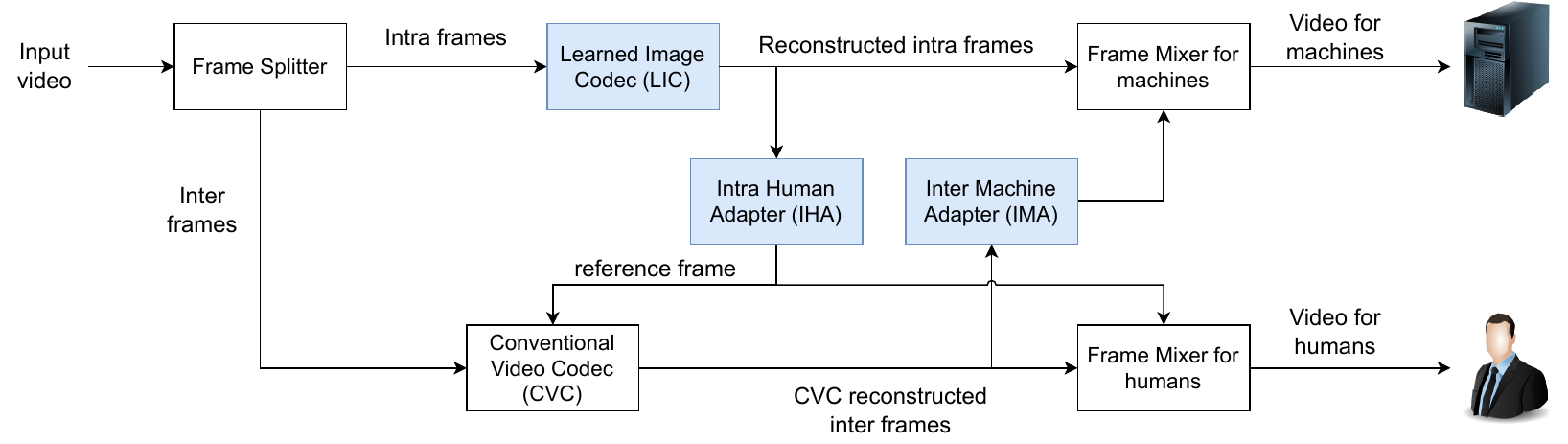}}
       
  \end{minipage}
  \centerline{(a) NN-VVC system overview}
  \hspace{1cm}
  
  \begin{minipage}[b]{\linewidth}
    \centering
      \centerline{\includegraphics[width=\linewidth]{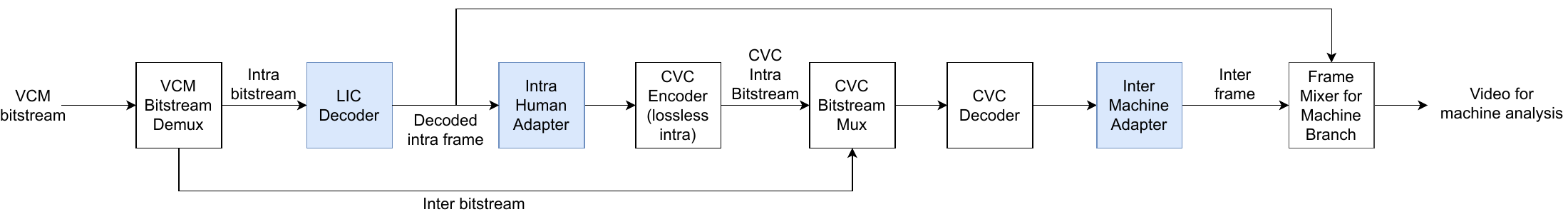}}
       
  \end{minipage}
  \centerline{(b) Decoding pipeline}
  
    \caption{The NN-VVC coding system, light blue color indicates a neural network component}
    \label{fig:nnvvc}
  
\end{figure*}

End-to-end learned video coding targeting human viewing has been the subject of intense research in recent years \cite{agustsson2020scale,aivc,mentzer2022vct, choiScalableVideoCoding2022}. However, despite the impressive progress over just a few years, end-to-end learned video codecs are still not able to outperform the latest traditional codecs, such as VVC \cite{cit:vvc}. On the other hand, end-to-end learned image codecs have been able to surpass the coding efficiency of these tools by large margins for both human consumption (in some specific settings, such as when using RGB color space and/or the multiscale structural similarity index quality metric) \cite{e2e_im_compression,scale_hyperprior,meanscale_hyperprior,minnen2018joint,mentzer2020highfidelity,clic_paper,cheng2020learned} and machine vision consumption \cite{icassp_paper, choiScalableImageCoding2022}. For this reason, we propose to harness the capability of Learned Image Codec (LIC) and the mature, widely adopted techniques of Conventional Video Codec (CVC) tools with a hybrid system that can deliver all-around higher coding performance for machine consumption against the state-of-the-art video codec VVC. We refer to our proposed hybrid codec as NN-VVC.

In NN-VVC, the LIC is used to perform intra-frame coding. For inter-frames, it takes advantage of the well-developed traditional coding tools of VVC, using the LIC-coded frames as the reference pictures. However, VVC may also be used to encode intra-frames in some cases (this is referred to as fallback mode, more information in the next sections). As shown in \autoref{fig:nnvvc}, at encoding time, the intra-frames are coded by the LIC. The Intra Human Adapter (IHA) then processes the LIC-reconstructed intra-frames to obtain filtered reconstructed intra-frames that are better suited as reference pictures for VVC (\autoref{ssec:iha}). The filtered reference frames are used by a CVC encoder to code the inter frames in a lossy fashion. Finally, the bitstream multiplexer (muxer) merges the intra-bitstream to the CVC bitstream, resulting in the VCM bitstream for transmission.

At decoding time, the VCM bitstream is decomposed to intra and inter-bitstreams using a bitstream demuxer. The intra-bitstreams are decoded by the LIC decoder, thus obtaining reconstructed intra-frames, which are then given as inputs to IHA. The outputs of IHA are used as reference frames to decode inter frames of the inter-bitstream. This can be achieved by modifying a standard CVC decoder to input decoded reference frames, which has the disadvantage that legacy CVC decoder implementations cannot be used as such. Another possibility, which is enabled by the lossless coding capability of state-of-the-art video codecs, such as VVC, is depicted in \autoref{fig:nnvvc} and enables the use of a CVC decoder without modifications. The outputs of IHA are losslessly encoded by a CVC encoder to produce the CVC intra-bitstream. From there, a CVC-compliant bitstream is obtained by multiplexing the CVC intra-bitstream and the inter-bitstream using a bitstream muxer. The CVC decoder then decodes the CVC-compliant bitstream. The output inter frames are further enhanced for task performance with an Inter Machine Adapter (IMA - \autoref{ssec:ima}). Finally, the video for machine consumption is formed based on the intra-frames and the enhanced inter-frames.

\subsection{Self-supervisedly Learned Image Codec (LIC)}
\label{ssec:lic}
\begin{figure*}[ht]
    \begin{minipage}[b]{0.95\linewidth}
    \centering
      \centerline{\includegraphics[width=\linewidth]{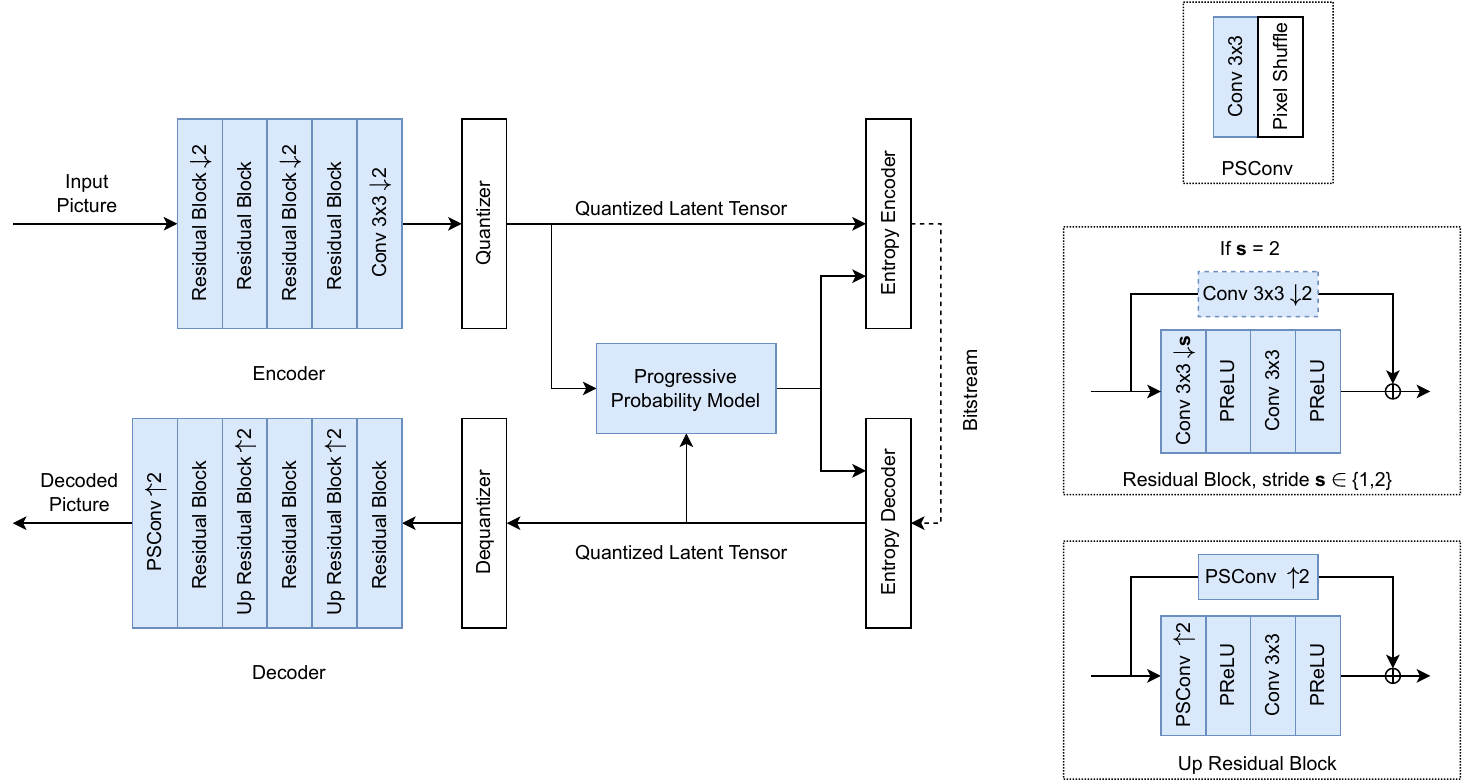}}
       
    \end{minipage}
    \caption{The learned image codec (LIC) for intra-frames.}
    \label{fig:lic}
\end{figure*}

The superiority of the ICM systems over VVC in \cite{icassp_paper, icme_paper, choiScalableImageCoding2022} motivates us to replace the intra coding in VVC with a learned ICM codec in order to get better machine task performance. We use the self-supervised image coding for machines system proposed in \cite{le2022_gans}, where the coding system is trained using a task network without annotations for the training data. More specifically, this system comprises a convolutional neural network (CNN) based encoder, a CNN-based decoder, a CNN-based probability model, and an Asymmetric Numeral Systems (ANS) entropy codec \cite{ans}. \autoref{fig:lic} shows an overview of the LIC. The input image $\img$ is transformed by the encoder $\Model{\litencoder}{}$ (parametrized by $\boldsymbol{\theta_E}$) to a latent tensor $\latent=\encoder{\img}$, then quantized and compressed to a bitstream by the entropy encoder. At decoding time, the entropy decoder decompresses the bitstream to the quantized latent tensor $\iqlatent$. Next, $\iqlatent$ is dequantized and restored to the image domain by the decoder $\resimg=\decoder{\iqlatent}$. The entropy coding process requires prior distributions of $\iqlatent$, which are provided by a progressive probability model proposed by Zhang \etal\cite{zhang2022pms}.
\newparagraph{Training method:}
We follow the same training strategy as proposed in \cite{icassp_paper} to obtain multiple image compression model checkpoints that achieve different qualities and bitrates. The LIC is trained to minimize three quantities: bitrate estimation, task loss, and distortion loss. The bitrate (or rate for simplicity) estimation is defined as the cross-entropy between the true distribution $q_{\iqlatent}$ of $\iqlatent$ and its estimation $p_{\iqlatent}$ made by the probability model:
\begin{equation}
   \lossrate = \mathbb{E}_{\iqlatent \sim q_{\iqlatent}} \left[-\log_{2} \prior{\iqlatent}(\iqlatent)\right] \\
\end{equation}
We employ the Mean-Squared Error (MSE) as the distortion loss to make the training more stable.
In order to train the LIC to be task-agnostic, we use the feature-domain, multi-layer distortion loss $\lossproxy$ proposed in \cite{le2022_gans} as a proxy for the real task loss. This training technique comes with important advantages that unlock the practicality of our method. Firstly, with a surrogate loss term, the optimization objectives are not tied to any particular vision task or network architecture, therefore the codec can have significantly better generalizability to different downstream tasks. Secondly, the training data is not constrained to the availability of annotations for multiple vision tasks, which is a requirement in training with task losses. This enables self-supervised learning of the codec on a large quantity of data. Lastly, when using the proxy loss it is easier to train the LIC with small patches of images instead of full images, leading to a significant reduction in computational resource consumption.
The final training objective is given as a linear combination of the loss terms:
\begin{equation}
   \Loss{total} = \wrate\lossrate + \wmse\lossmse + \wtask\losstask
\end{equation}
where $\wrate, \wmse, \wtask$ are scalar numbers whose values are decided by functions of epoch number, described as Loss Weighting Strategy (LWS) \cite{icassp_paper} in \autoref{eq:lossstrat}. By using LWS we are able to obtain model checkpoints that offer a wide range of output bitrates.

\subsection{Intra human adapter (IHA)}
\label{ssec:iha}
Especially on lower bitrates, the reconstructed intra frames coded with the LIC may contain different types of artefacts, such as the checkerboard artefacts that can be seen in \autoref{fig:subjective_comp} and were studied in \cite{blurpool_zhang2019making,le2022_gans}. While these artefacts do not affect the machine task performance when coding images, they might cause a significant degradation of compression efficiency for the CVC, as the LIC-reconstructed intra-frames are used as reference frames for inter-frame prediction. To remove these artefacts, we use IHA to enhance the LIC-reconstructed intra-frames in terms of the peak signal-to-noise ratio with respect to the corresponding uncompressed intra-frames. IHA is formulated as $H$ in $\resimg_H=H(\resimg$),  where $\resimg$ and $\resimg_H$ are the LIC reconstructed image and Intra Human Adapted image, respectively. The structure of the IHA is based on the enhancement filter structure proposed in \cite{pfilter_ahonen2021}, which is essentially a convolutional autoencoder with skip connections. The differences with respect to \cite{pfilter_ahonen2021} consist of an extra skip connection from input to output tensor and combined Quantization Parameter (QP) and resolution injection blocks before every up- and downsampling convolutional layer. Injection blocks concatenate the QP and resolution information of the processed frames together and feed them to a simple linear layer followed by a parametric rectified linear unit (PReLU). After this, the output is repeated to match the size of the filter's features where the injection is performed, to which it is then concatenated.

\subsection{Inter machine adapter (IMA)}
\label{ssec:ima}
In order to adapt the CVC reconstructed inter frames $\resimg_{cvc}$ to perform better on machine tasks, we use the IMA, formulated as $M$ in $\resimg_M=M(\resimg_{cvc}$), where $\resimg_M$ is the machine adapted inter frame. The structure of the IMA is similar to that of the IHA, except that it does not contain the QP and resolution injections which, based on empirical evaluation, did not bring any benefits to the IMA.

\subsection{Fallback mode and spatial re-sampling}
\label{ssec:fback}
When coding a video with an extremely low LIC quality, even the IHA cannot suppress the LIC artefacts well enough for the CVC compression to remain efficient. To overcome this problem, we introduce the fallback mode, which is activated when a certain threshold is reached for the expected quality of LIC. In fallback mode, the whole LIC branch including the IHA is switched off and only the CVC is used to code the video (including the intra frames). The CVC by itself is able to handle the low bitrate coding efficiently and by adapting both intra- and inter-frames with fallback mode designated IMA (F-IMA), the machine task performance of the reconstructed video will be increased over the plain CVC. The structure of the F-IMA is equivalent to that of the IHA.

Since the LIC, IHA, and IMA are all trained with images having resolutions less than 1920$\times$1080, to efficiently handle data that has a higher resolution, we apply a simple spatial down-sampling to input images/videos that have a resolution higher than 1920$\times$1080, by using a downsampling factor of 3/4. Then, the reconstructed output of the LIC decoder and IMA are upsampled by a factor of 4/3 to restore the original resolution. Another possible option, that might be part of our future work, would be to expand the training data to include images up to 4K and 8K to achieve better performance on higher-resolution images compared to spatial resampling.

\subsection{Adapter training}
To train the different types of adapters (namely IHA, IMA, F-IMA), for each adapter type, we use the following training  loss with different proxy loss weights $\wproxyA$:
\begin{align}
   \Loss{total_A} &= \Loss{mse_A} + \wproxyA\Loss{proxy_A} \\
\Loss{mse_A}&=\frac{1}{N}\sum_{i = 1}^{N}\lVert\img_1 - \resimg_2\rVert^2
   \label{eq:adapterloss}
\end{align}
where $\img_1$ is the uncompressed frame, and $\resimg_{2}$ is the adapted output frame. For the IHA, we set the $\wproxyA$ as 0, resulting in the use of only the $\Loss{mse_A}$. For both IMA and F-IMA, we set a positive
scalar as $\wproxyA$, which enables the use of the $\Loss{proxy_A}$ in
training. $\Loss{proxy_A}$ is defined as a proxy loss similar to the LIC training, but the backbone part of the $\textit{maskrcnn\_resnet50\_fpn}$\footnote{The pre-trained models can be found at \url{https://pytorch.org/docs/stable/torchvision/models.html}} by torchvision \cite{torchvision} is used to extract the features from the adapted and uncompressed frames.

\subsection{Bitrate control mechanism} 
\label{ssec:rate-control}
One of the most important aspects of any video codec is the ability to control the size of the produced bitstream and the corresponding output quality. We use a couple of different techniques to achieve an efficient bitrate control mechanism. Several LIC models were trained to achieve different intra-frame bitrates as described in \autoref{ssec:lic}. Specifically, six different LIC models were trained to achieve similar bitrates as CVC when configured to operate with six different QPs in the simulation conditions. When an intra-frame is to be coded to a certain target bitrate, we select the LIC model that achieves the closest bitrate to the target bitrate, based on a look-up table. For example, if we want to code a certain intra-frame to produce a bitstream size close to the bitstream size of a CVC-encoded intra-frame with QP 33, and the closest model of LIC corresponds to QP 32, it is then used to perform the encoding. When performing video coding, we set a target $QP_{inter} \in [0, 63]$ for CVC to code the inter-frames, while the intra-frames are coded with an -5 offset: $QP_{intra} = QP_{inter} -5$. If more compression is needed, the QP of CVC can be increased.
\subsection{Bit-exact reconstruction}
\label{ssec:bitexact}

By default, convolutional layers in common NN-based systems operate in the floating point domain. When executed in different computing environments, the results from these operations may be different. In critical situations, such as for the components of the probability model, the discrepancies lead to a total corruption of the decoded data.
Therefore, to make a codec useful in the real world, it is crucial to make sure some operations produce the same results regardless of the processing environments.
In order to achieve bit-exactness in different computing environments, we perform the convolutional operations of critical components in the quantized domain, as proposed in \cite{intconv_anonymous_submitted}.
It is required that the quantized convolutions are applied in the decoder and probability model of LIC, and IHA. Quantized convolutions are optional for IMA and F-IMA, but when applied, the decoding results will be deterministic in different environments, with negligible effects on the coding performance. All the experimental results that are shown in \autoref{ssec:ex_results} use the quantized convolutions only on LIC and IHA.
\section{Experiments}
\label{sec:experiments}
\subsection{Model training}
\label{ssec:ex_setup}
\newparagraph{LIC training:} We applied similar training techniques with the Loss Weighting Strategy (LWS) as proposed in \cite{icassp_paper}. On top of that, in order to get a better starting point, we trained the LIC model in the first phase with a weighted summation of $\lossrate$ and $\lossmse$. The training data preparation for the first phase was the same procedure described in \cite{zhang2022pms}, which uses a random subset of 340K images from the training set of Open Images V6 \cite{openimages}. The training data for the following phases is a random subset of 6K images from the Open Images V6 train set every epoch. Additionally, we reduced the training time by using approximately half the number of epochs for the phases after warming up, which are specified by $p_2, p_3, p_4$ \cite{feature_residuals_seppala2021}, compared to \cite{icassp_paper}. The final LWS formulation is specified by functions of the epoch number $n$:

\begin{equation}
  \begin{split}
    \wmse&=1, \\
    \wtask&=\begin{cases}
      0,               & n < p_1   \\
      4\boldsymbol{\psi}(n-p_1,1.01), & n \ge p_1
    \end{cases}, \\
    \wrate&=\begin{cases}
      0.01,                  & n < p_1         \\
      0,                  & n < p_2         \\
      2\boldsymbol{\psi}(n-p_2,1.01),    & p_2 \le n < p_3  \\
      c,                  & p_3 \le n < p_4 \\
      c+2\boldsymbol{\psi}(n-p_4,1.02), & n \ge p_4      \\
    \end{cases},
  \end{split}
  \label{eq:lossstrat}
\end{equation}
where $\boldsymbol{\psi}(x, y) = 10^{-3}(y^{x} - 1), p_1 = 50, p_2 = 62, p_3=85,p_4=107$ and $c=2\boldsymbol{\psi}(p_3-p_2-1,1.01)$.
We collected the 6 model checkpoints that offer close average bitrates to QPs [22, 27, 32, 37, 42, 47] of VTM 12.0 \cite{vtm} (reference software of VVC). 
The collected checkpoints correspond to epoch numbers $n$ [68, 80, 170, 220, 270, 320], respectively. Models trained for a small number of epochs were not found to be sufficiently optimized, thus we performed an additional finetuning process of the checkpoint corresponding to epoch 68 (QP 22) for another 50 epochs with the same training settings except for fixed loss weights which could be obtained with \autoref{eq:lossstrat} where $n=68$.
\newparagraph{IHA training:} 30K images were randomly selected from the training split of Open Images, and encoded and decoded by using all the available LIC models. Multiple random patches of size $256\times256$ were cropped out from each reconstructed image and the IHA was trained for 88 epochs with a batch size of 50.
\newparagraph{IMA training:} we used the training split of the BVI-DVC dataset \cite{ma2021bvidvc} to generate the training data, which was obtained by running the NN-VVC system for target QPs [22, 27, 32, 37, 42, 47, 52] with IMA turned off. The IMA model was trained for 50 epochs with patches of size $240\times240$ extracted similarly to IHA training. The training data for the F-IMA was generated by coding and reconstructing the training split of BVI-DVC using only the CVC with QPs [22, 27, 32, 37, 42, 47, 52, 57, 62]. Even though the fallback mode was used only in cases where LIC QP $>$ 49, it was empirically noted that including the training data generated with lower QPs makes the model more robust when used in the fallback mode. F-IMA was trained for 28 epochs with batches of 96 patches of size $256\times256$. Following the total loss defined in \autoref{eq:adapterloss}, proxy loss weight $\wproxyA$ was set to 0, 0.1 and 0.015 for IHA, IMA and F-IMA, respectively. Adam optimizer \cite{adam} with a learning rate of $2\expnum{-4}$ was used for every adapter.

\newcommand{\width}{0.43}
\begin{figure*}[ht]
  \centering
  \begin{minipage}[t]{\width\linewidth}
    \centering
    \centerline{\includegraphics[width=\linewidth]{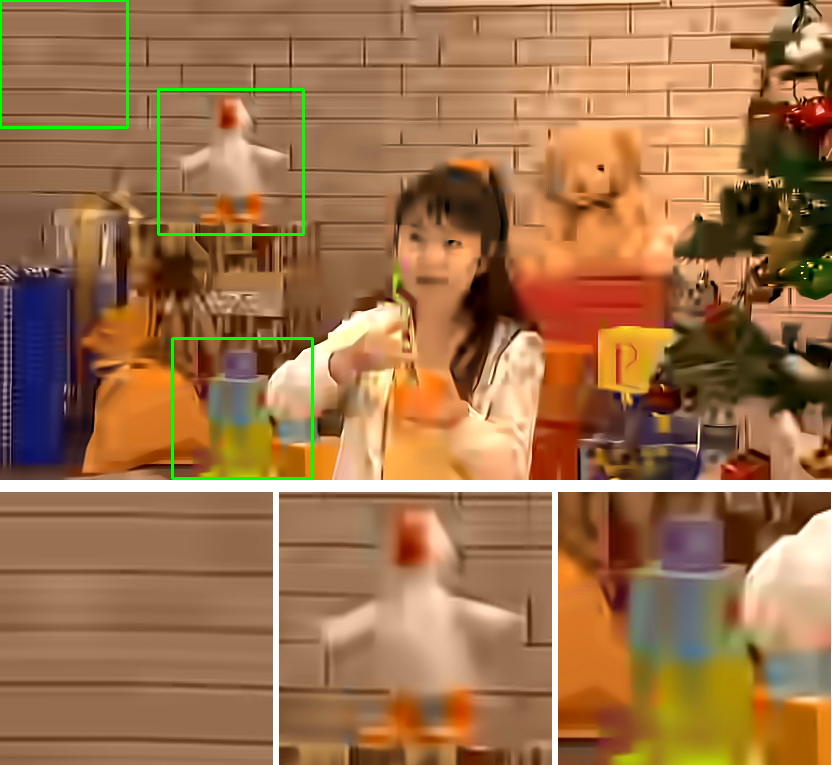}}
    \centerline{VVC, frame 0 (intra frame)}
  \end{minipage}
  \hspace{0.5cm}
  \begin{minipage}[t]{\width\linewidth}
    \centering
    \centerline{\includegraphics[width=\linewidth]{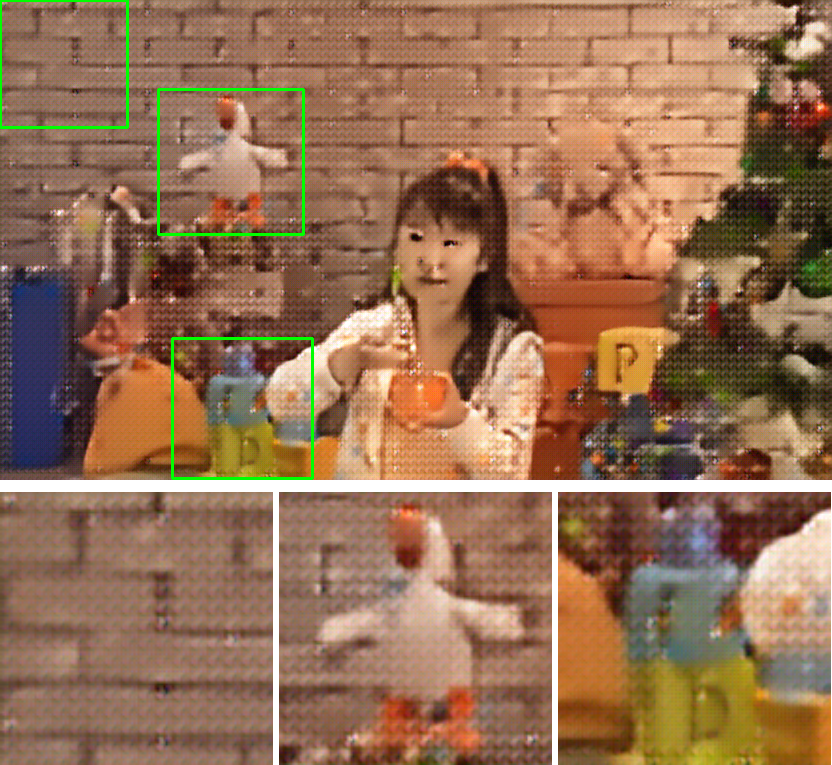}}
    \centerline{NN-VVC, frame 0 (intra frame)}
  \end{minipage}
  \vspace{0.2cm}
  
  \begin{minipage}[t]{\width\linewidth}
    \centering
    \centerline{\includegraphics[width=\linewidth]{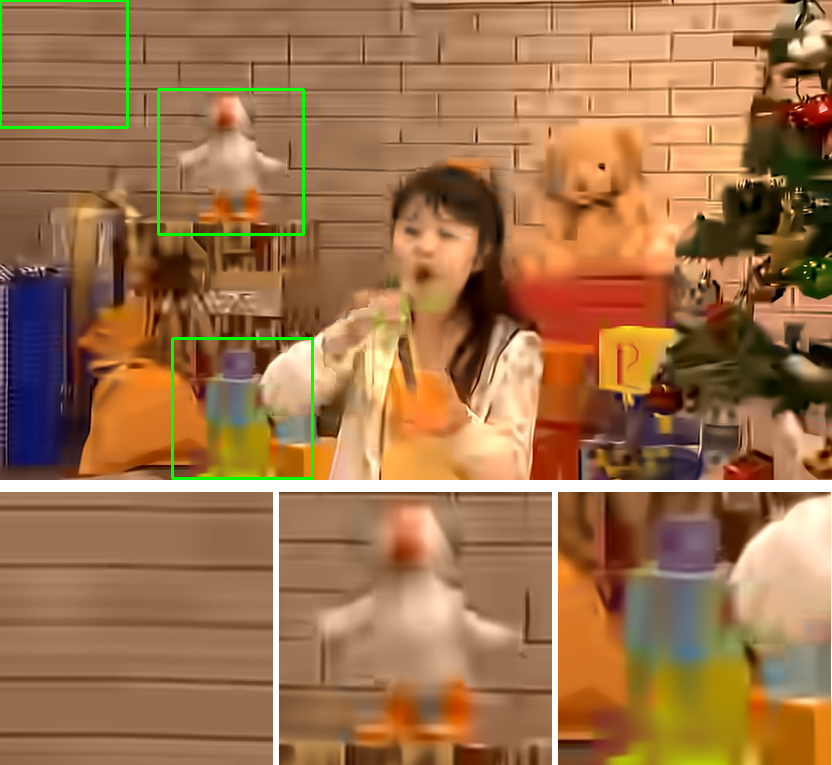}}
    \centerline{VVC, frame 7 (inter frame)}
  \end{minipage}
  \hspace{0.5cm}
  \begin{minipage}[t]{\width\linewidth}
    \centering
    \centerline{\includegraphics[width=\linewidth]{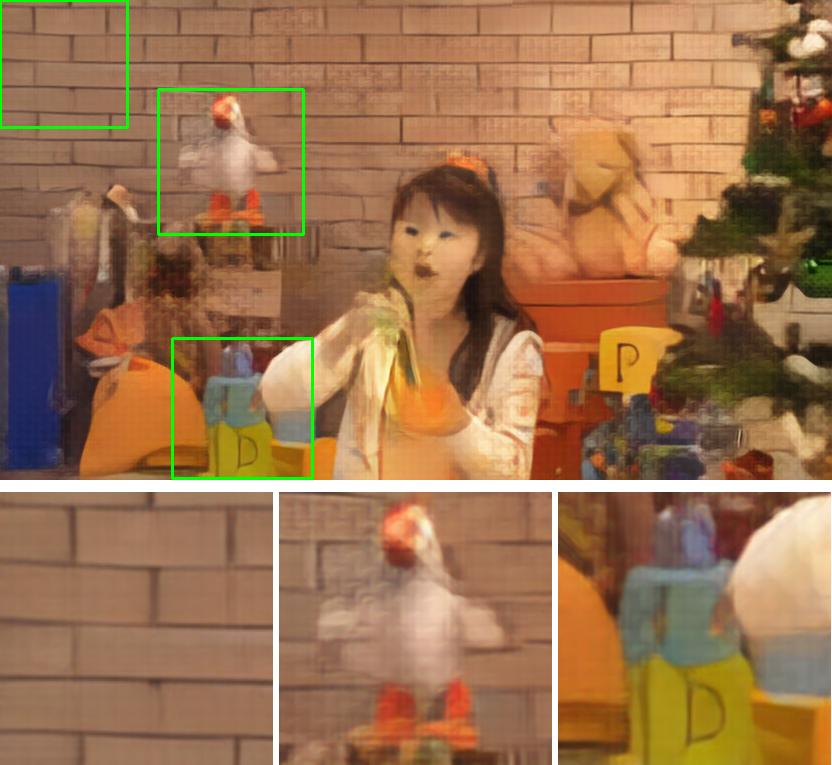}}
    \centerline{NN-VVC, frame 7 (inter frame)}
  \end{minipage}
  \caption{The output from VVC compared to our proposed codec. The input sequence \texttt{PartyScene\_832x480\_50\_val} was coded with \texttt{IntraPeriod = 32, QP=52}. Our codec at the same bitrate managed to preserve details of the foreground objects better. The strong edges in the background are also more visible.}
  \label{fig:subjective_comp}
\end{figure*}
\newcommand{\rdcurvwidth}{0.33}
\begin{figure*}[ht]
  \begin{minipage}[b]{\rdcurvwidth\linewidth}
    \centering
      \centerline{\includegraphics[width=\linewidth]{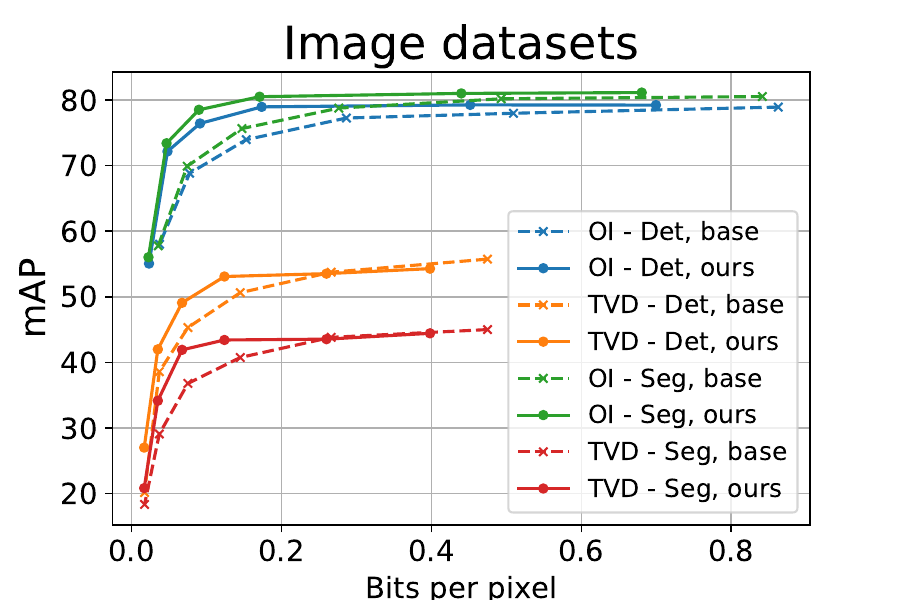}}
  \end{minipage}%
  \begin{minipage}[b]{\rdcurvwidth\linewidth}
    \centering
      \centerline{\includegraphics[width=\linewidth]{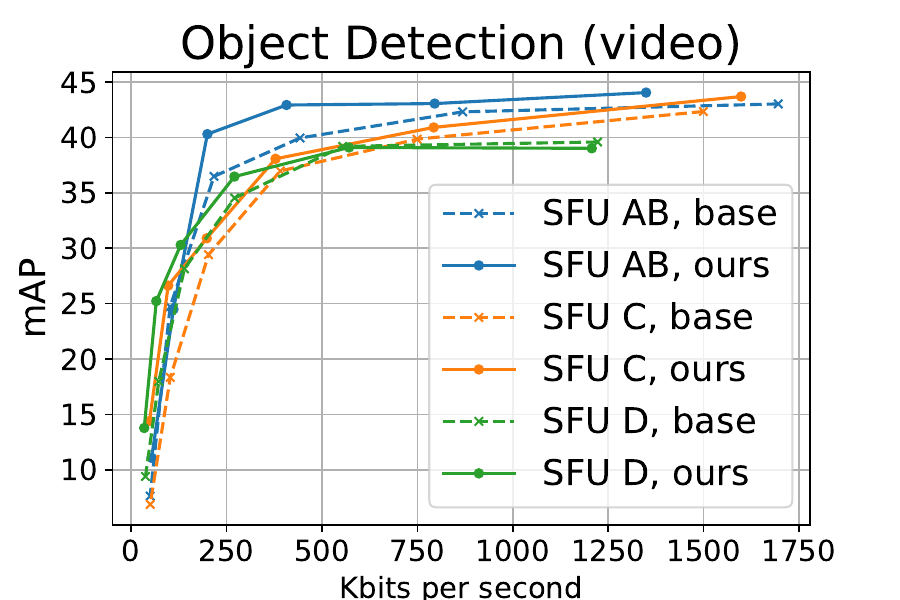}}
  \end{minipage}%
  \begin{minipage}[b]{\rdcurvwidth\linewidth}
    \centering
    \centerline{\includegraphics[width=\linewidth]{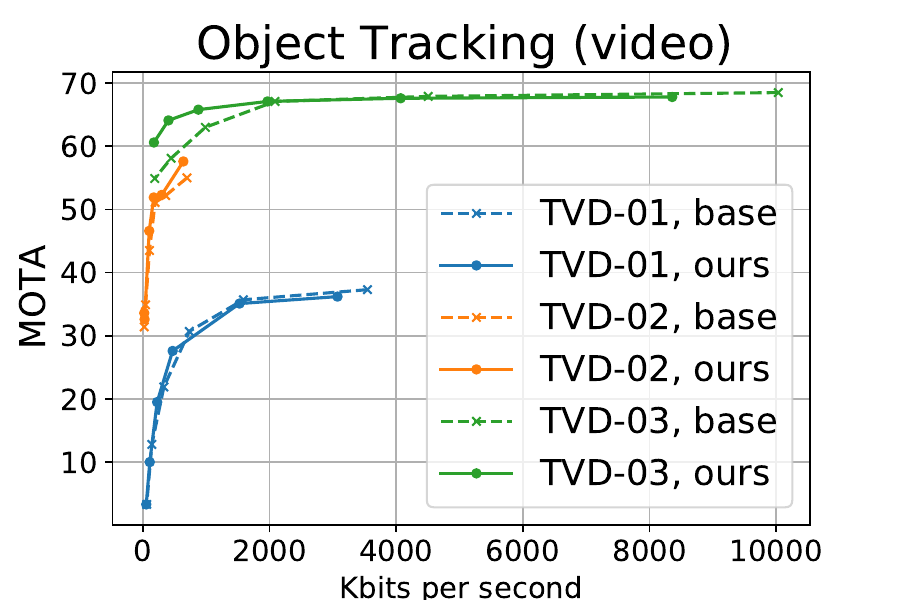}}
  \end{minipage}%

\caption[]{Evaluation results across multiple vision tasks (Object detection, instance segmentation and multiple object tracking) on multiple datasets (Open Images, TVD image, TVD video and SFU video), in comparison to VVC as the baseline.}
\label{fig:rdcurves}
\end{figure*}

\renewcommand\theadfont{\bfseries}
\begin{table*}[ht]
   \caption{Codec performance results compared to VVC/H.266 on different vision tasks. The results are evaluated using the Bj{\o}ntegaard Delta \cite{bdrate} against rate (BD-rate) and task performance metric (BD-task). The scores are presented in ``BD-rate $|$ BD-task'' format, where BD-task represents BD-MOTA for TVD dataset and BD-mAP for the rest of the datasets.}
   \label{tab:main_results}
   \centering
      \begin{tabular}{@{}clccccc@{}}
         \toprule
                                                                        &                              & \multirow{2}{*}{\thead{Object detection}} & \multirow{2}{*}{\thead{Instance segmentation}} & \multirow{2}{*}{\thead{Object tracking}} & \multicolumn{2}{c}{\thead{Runtime ratio to VVC}}\\
                                                                        
                                                                        &&&&&{Encoding} &{Decoding} \\\midrule
         \rule{0pt}{12pt}
         \multirow{2}{*}{\rotatebox[origin=c]{90}{Image}} & Open Images\cite{openimages} & -53.04 \% $|$ 4.64(*)  & -51.76 \% $|$ 4.84~~~~ & --   & 0.09	& 26.63
         \\
         \rule{0pt}{12pt}                                                
         & TVD image\cite{gaoTVD2022}   & -30.04 \% $|$ 3.32~~~~ & -38.07 \% $|$ 3.60~~~~ & --   & 0.09	& 19.92         \\ 
         \midrule
         \multirow{6}{*}{\rotatebox[origin=c]{90}{Video}} 
                         & SFU AB\cite{choi2020SFUDet}                       & -13.94 \% $|$ 1.48~~~~ & --                     & --      & 0.53	& 16.75 \\
                                                                        & SFU C\cite{choi2020SFUDet}                         & -32.76 \% $|$ 3.66~~~~ & --                     & --  &                   0.39	& 21.39
                                                                        \\
                                                                        & SFU D\cite{choi2020SFUDet}                         & -34.55 \% $|$ 3.07(*)  & --                     & --  &                   0.37	& 37.58
                                                                        \\
                                                                        & TVD-01\cite{gaoTVD2022}      & --                     & --                     & -7.84 \% $|$ 0.56~~    
                                                                        & 0.27	& 33.12
                                                                        \\
                                                                        & TVD-02\cite{gaoTVD2022}      & --                     & --                     & -14.31 \% $|$ 1.28(*)  &
                                                                        0.24	& 31.76
                                                                        \\
                                                                        & TVD-03\cite{gaoTVD2022}      & --                     & --                     & -57.38 \% $|$ 2.67~~~~ &
                                                                        0.37	& 27.10
                                                                        \\ \bottomrule
      \\   
      \end{tabular}%
\\
\footnotesize{($\ast$) marks the cases where the task performance scores of certain proposed datapoints are lowered to have a monotonic rate-performance curve in order to make BD-rate calculation possible. BD-task is calculated with original values.}
\end{table*}

\subsection{Evaluation setup and results}
\label{ssec:ex_results}
\begin{figure*}[ht]
    \centering
    \begin{minipage}[b]{0.48\linewidth}
    \centering
    \centerline{\includegraphics[width=\linewidth]{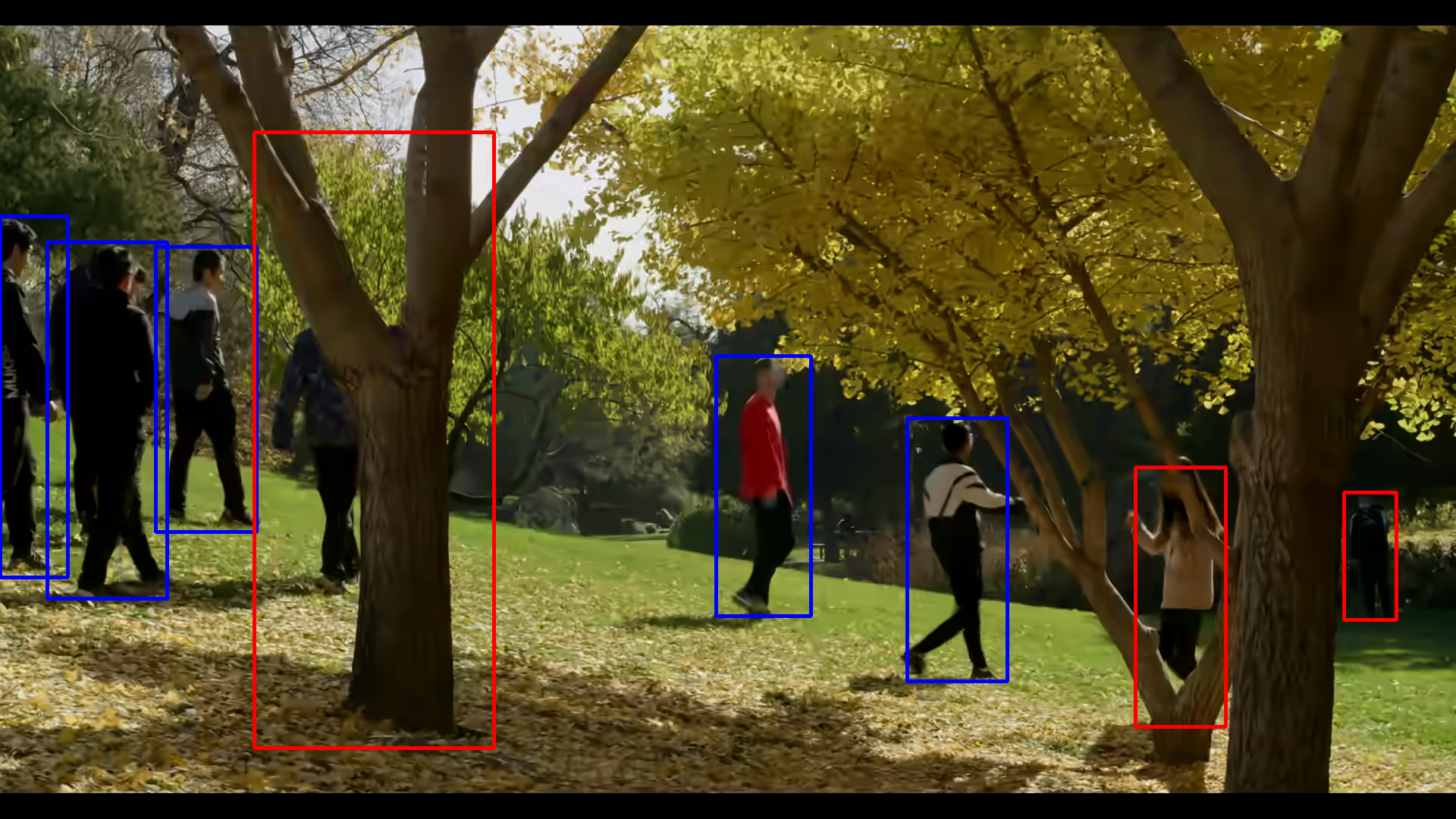}}
    \centerline{VVC QP42 frame 4}      
    \end{minipage}
    \begin{minipage}[b]{0.48\linewidth}
    \centering
    \centerline{\includegraphics[width=\linewidth]{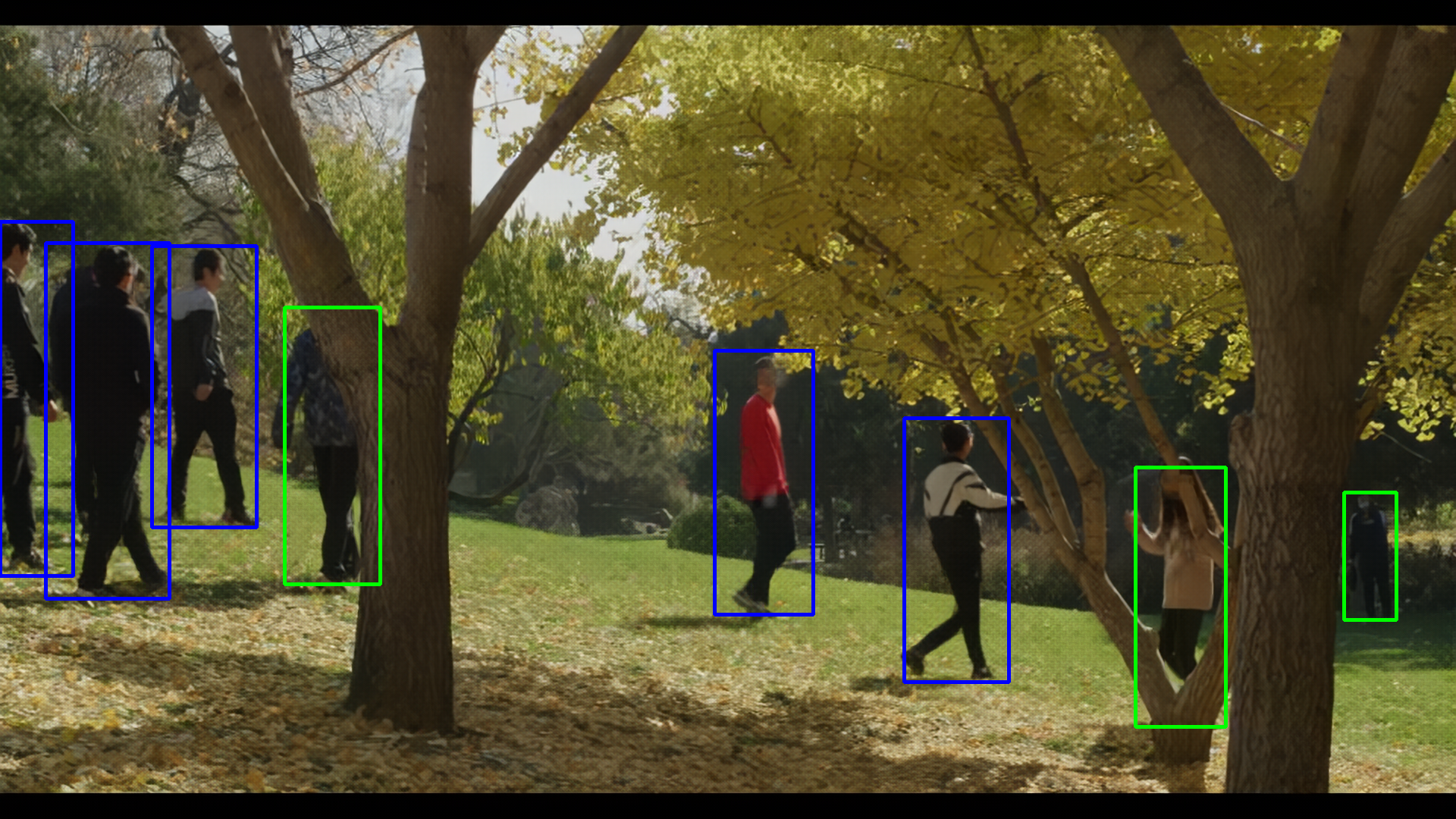}}
    \centerline{NN-VVC QP42 frame 4}      
    \end{minipage}
  \begin{minipage}[b]{0.48\linewidth}
    \centering
    \centerline{\includegraphics[width=\linewidth]{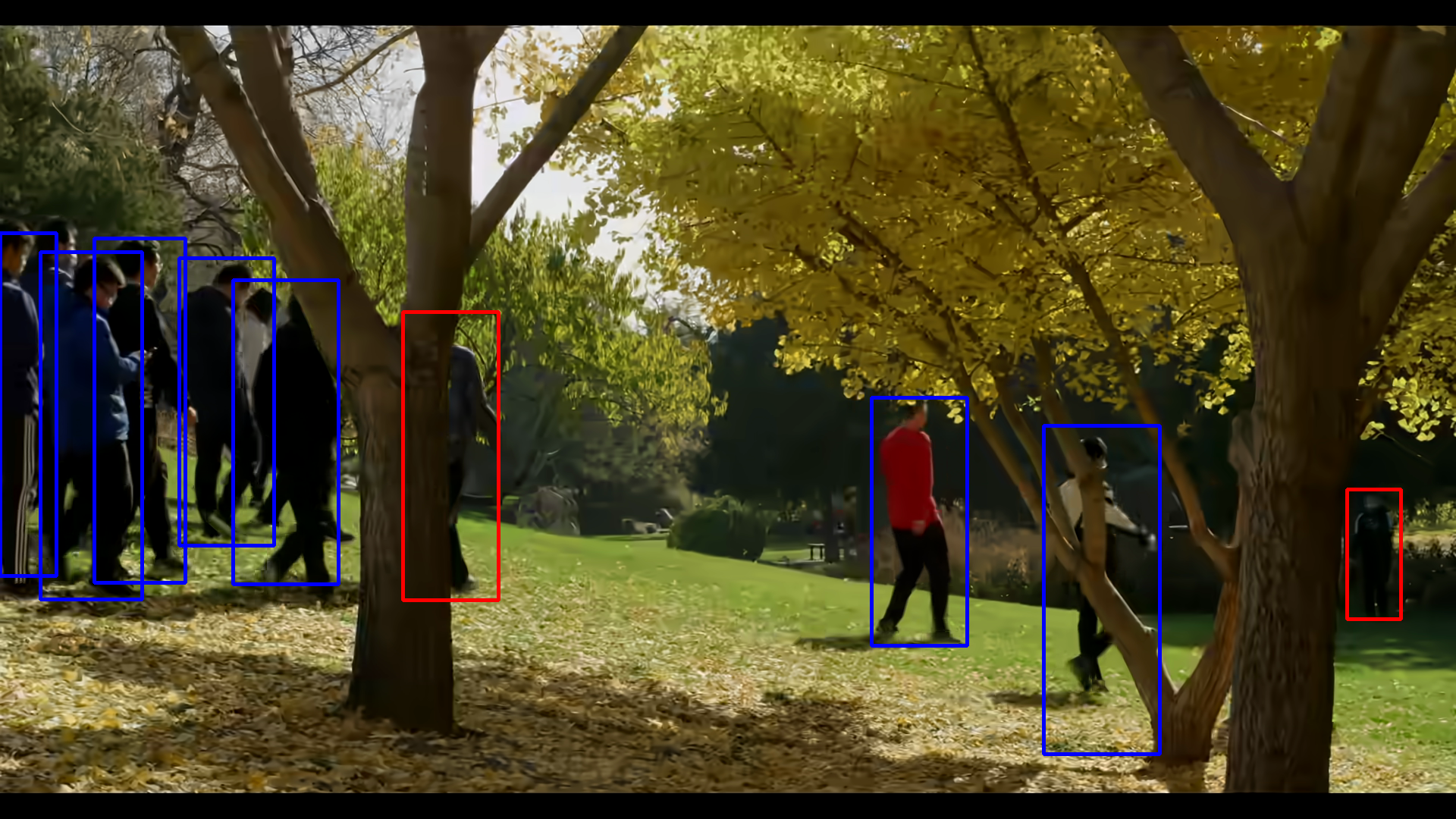}}
  \centerline{VVC QP42 frame 75}     
  \end{minipage}
  \begin{minipage}[b]{0.48\linewidth}
    \centering
    \centerline{\includegraphics[width=\linewidth]{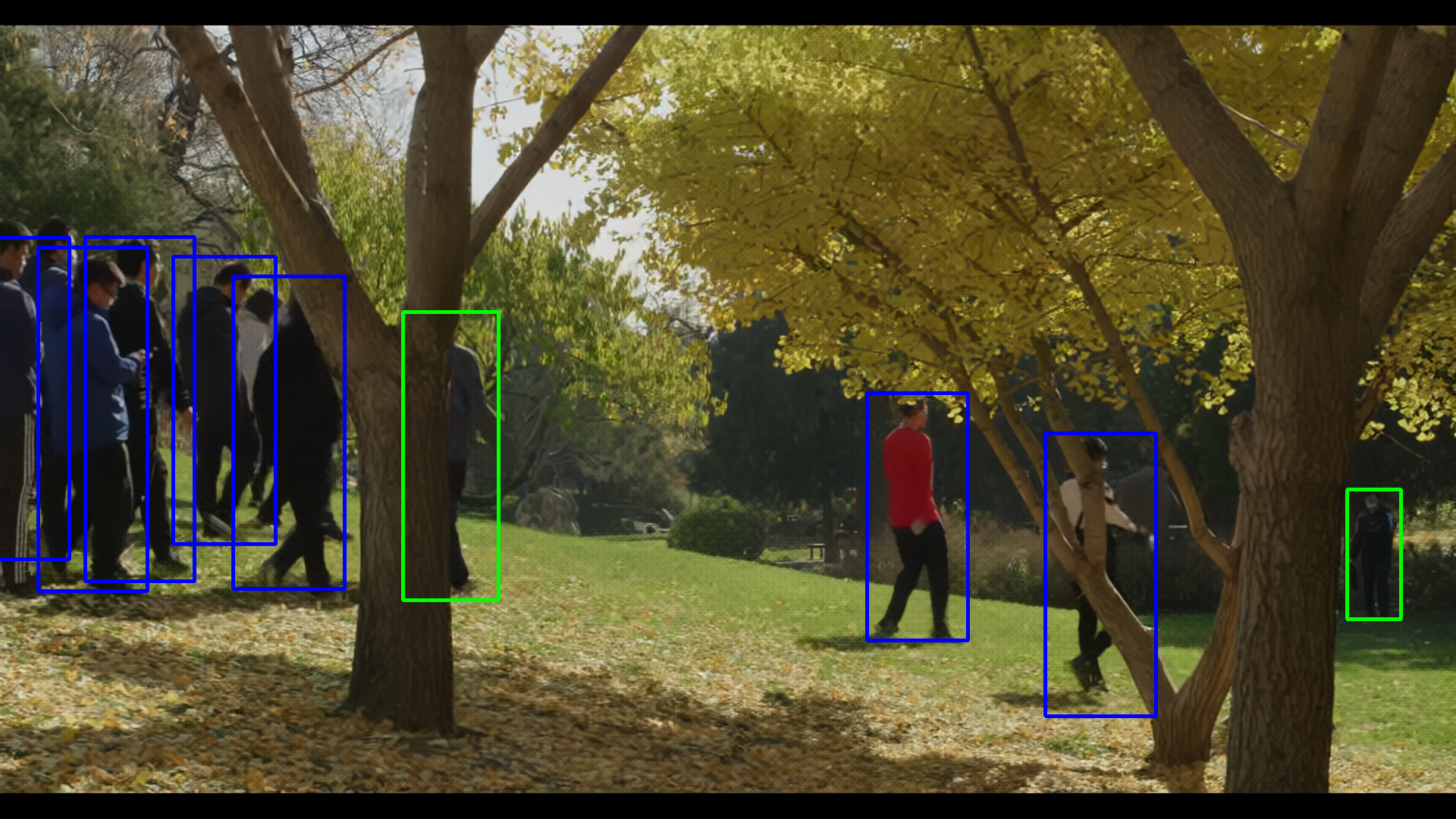}}
  \centerline{NN-VVC QP42 frame 75}      
  \end{minipage}
  
\caption[]{Comparison of the predicted bounding boxes from VVC and NN-VVC reconstructed inter frames on TVD-03 video. Blue, red and green bounding boxes represent common correct predictions, missing/incorrect predictions only on VVC reconstructed frames, and correct predictions only on NN-VVC 
reconstructed frames, respectively. Note that all the red bounding boxes are missing predictions, except the largest one in the top left image.
}
\label{fig:bbox_comparison}
\end{figure*}

We used the same environment for all the benchmarks and evaluations in this work. Our testing hardware was an NVIDIA DGX1 machine with 8 Tesla V100 GPUs and an 80-threaded Intel Xeon CPU E5-2698 v4 CPU.
We evaluated our method based on the Common Test Conditions for evaluating the Call for Proposal (CfP) responses, issued by the MPEG VCM group \cite{vcm_ctc}. Following this evaluation framework, the performance of the codecs was measured for three vision tasks, on two image datasets, one of which is developed for MPEG VCM activities based on Open Images V6 \cite{openimages}, and 17 video sequences (3 from TVD dataset \cite{gaoTVD2022} and 14 from SFU dataset\cite{choi2020SFUDet}). The video sequences were categorized into classes for class-wise performance. \autoref{tab:main_results} reports the performance of our codec in terms of BD-rate and BD-\textit{task} \cite{bdrate} calculated from 6 different rate points against the VVC anchor as the performance metric. The BD-mAP and BD-MOTA indicate the mean average precision (mAP \cite{mAP_coco}) gain and the multiple object tracking accuracy (MOTA \cite{milan2016mot16}) gain, respectively, at an equivalent bitrate. We use BD-\textit{task} to refer collectively to BD-mAP and BD-MOTA. Ideally, a complete comparison would consider also other machine-oriented video codecs in addition to the state-of-the-art conventional video codec VVC. However, the codecs in \cite{vcm_feature_based, huangHMFVCHumanMachineFriendly2022, choiScalableVideoCoding2022} are ``feature compression'' methods that partially employ the task networks in their pipeline. Additionally, they were not evaluated following the MPEG common test conditions for video coding for machines (VCM CTC). In these works, the evaluation datasets and QPs differ from each other as well as from the VCM CTC. In our evaluation, we followed the VCM CTC as it represents a structured and extensive approach for evaluating codecs that target machine analysis. In addition, this enables future works that follow VCM CTC to be easily compared to ours. 
\newparagraph{Visual quality:} Targeting vision task performance over pixel-fidelity, our codec seeks to conserve the semantic features of the input. In high bitrates, both these features and pixel fidelity can be preserved at the same time. In order to demonstrate the differences in bit-allocation priority of our codec, we deliberately coded the input sequence in a low bitrate setting (QP 52), in comparison to VVC. The outputs can be seen in \autoref{fig:subjective_comp}. The intra-frames show the coding artifact patterns due to the low bit budget. At this bitrate, VVC intra codec suffered from traditional coding artifacts, whereas our LIC codec suffered from the ``checkerboard'' patterns which are commonly found in CNN-based image codecs. The IHA then heavily attenuated the patterns, making the intra-frame a more appropriate reference for inter-coding. The investigated inter-frames in the figure depict how these patterns propagated to inter-frames coded by CVC. Note that the IMA might also introduce these checkerboard artifacts to the inter-frames. Besides the differences in coding artifact patterns, compared to the output inter-frame from VVC, more edges, better-defined shapes, and well-preserved texts in the foreground objects can be observed in the output of our NN-VVC codec. These features are critical information to many vision tasks.
\newparagraph{Task performance benchmark:} Our codec outperformed VVC by a significant margin. On average, NN-VVC achieved a BD-task gain of 4.1 and 2.12 over the anchor for the tested image and video datasets, respectively. Corresponding average BD-rate reductions were \(-43.20\%\) for images and \(-26.8\%\) for videos. \autoref{fig:rdcurves} shows the rate-distortion curves from where the BD metrics were calculated. The most significant part of the gains over VVC came from the lower half of the bitrate range.
\autoref{fig:bbox_comparison} shows two examples of the prediction accuracy gain on TVD-03 object tracking sequence by comparing the bounding boxes detected from the inter frames reconstructed by VVC and NN-VVC. Both examples illustrated that, because of the heavy compression, the task network had difficulties predicting some of the bounding boxes for the VVC reconstructed frames. This is especially noticeable with instances that are harder to predict, such as the person further on the background in the right side of the frame, as well as most of the persons partly occluded by a tree. However, for the NN-VVC reconstructed frames, the task network was able to predict correctly as illustrated with the green bounding boxes.
\begin{table}[ht]
   \caption{System complexity measured by the number of multiply–accumulate (MAC) operations per pixel.}
   \label{tab:complexity}
   \centering

\begin{tabular}{@{}lcc@{}}
\toprule
\thead{Process}             & \thead{Complexity} & \thead{Number of  parameters} \\ 
                    &       (kMACs/pixel)                   & \\ \midrule
Intra encoding      & 1631.31                  & 4.3M                  \\
Intra decoding      & 1709.06                  & 6.5M                  \\
IHA                 & 163.62                   & 792K                  \\
IMA                 & 161.68                   & 782K                  \\
IMA - fallback mode & 163.22                   & 792K                  \\ \bottomrule
\end{tabular}
\end{table}
\newparagraph{Complexity and coding runtime:} \autoref{tab:complexity} shows the complexity of every NN-based component in our system. Similar to most of the NN-based image codecs, the design of our intra codec (LIC) distributes the computation between the encoder and decoder fairly evenly, unlike traditional codecs such as VVC which are designed and heavily optimized for the shortest decoding runtime. As a result, compared to VVC when tested on the same hardware and configurations as previously described for the evaluation environment, our encoder could be 2 to 10 times faster, while the decoder was 17 to 38 times slower in image and video coding as shown in \autoref{tab:main_results}. 
On the other hand, the decoding time of our system was \(2-19\%\) and \(79-92\%\) of the encoding time in the aforementioned video and image coding tests, respectively, even though the LIC decoder has a higher complexity than the LIC encoder. This was because of two main reasons: i) the inheritance of the VVC decoder in our codec and ii) the use of the progressive probability model \cite{zhang2022pms} in our LIC, which enabled a parallel decoding process of the intra-frames.

\subsection{Ablation study}
\label{ssec:abl_study}
In addition to the main results, an ablation study was conducted to show the importance of each main component of the NN-VVC system for machine task performance. Specifically, LIC, IHA and IMA were tested for coding videos. \autoref{tab:abl_res} contains BD-rate and BD-mAP/BD-MOTA results for SFU object detection \cite{choi2020SFUDet} and TVD object tracking \cite{gaoTVD2022} with different configurations. Note that the IMA in the table implies that either IMA or F-IMA was applied.

Starting from the base configuration ("No adapters"), where only LIC + CVC was used, we note that the LIC was able to introduce characteristics important to the machine tasks, especially with the SFU C and TVD-03. The IMA improved the performance significantly, except for SFU AB, because the LIC-coded intra images had introduced too much distortion to the reference images of the CVC, and eventually to the pictures that IMA is trying to adapt, which might introduce even more distortions in some cases. While the IHA itself did not generally improve the machine task performance as it was optimized only for MSE, its importance can be seen when being used together with the IMA and compared to the configuration where only IMA was used. Especially the SFU class AB, which is a problematic class for the IMA without IHA, was improved significantly. A general conclusion from this ablation study is that while some of the components worked better than others by themselves, they complemented each other and should be used together for achieving the best machine task performance gains.

\begin{table*}[ht]
\caption{Ablation study of machine task performances with different system configurations. BD-rate was used as a performance metric for object detection and object tracking. BD-mAP and BD-MOTA were used for object detection and object tracking, respectively.}
\label{tab:abl_res}
\centering
\begin{tabular}{cc|ccc|ccc|c}
\multirow{2}{*}{\textbf{Metric}}                                             & \multirow{2}{*}{\textbf{Configuration}} & \multicolumn{3}{c|}{\textbf{Object detection}} & \multicolumn{3}{c|}{\textbf{Object tracking}} & \multirow{2}{*}{\textbf{Average}} \\
                                                                             &                                         & SFU AB        & SFU C         & SFU D          & TVD-01       & TVD-02         & TVD-03        &                                   \\ \hline
\multirow{4}{*}{BD-rate $\downarrow$}                                                     & No adapters                             & 2.82 \%       & -25.50 \%     & -4.32 \%       & 8.84 \%      & 9.35 \%        & -14.94 \%     & -3.96 \%                          \\
                                                                             & IHA                                     & 2.04 \%       & -21.59 \%     & -2.21 \%       & 8.83 \%      & 17.70 \%       & -30.45 \%     & -4.28 \%                          \\
                                                                             & IMA                                     & 1.22 \%       & -31.02 \%     & -31.08 \%      & -8.95 \%     & -16.54 \%*     & -58.21 \%     & -24.10 \%                         \\
                                                                             & IHA + IMA                               & -13.94 \%     & -32.76 \%     & -34.55 \%*     & -7.84 \%     & -14.31 \%*     & -57.38 \%     & -26.80 \%                         \\ \hline
\multirow{4}{*}{\begin{tabular}[c]{@{}c@{}}BD-mAP $\uparrow$\\  BD-MOTA $\uparrow$\end{tabular}} & No adapters                             & -0.36         & 2.87          & 0.41           & -0.87        & -0.59          & -0.05         & 0.23                              \\
                                                                             & IHA                                     & -0.24         & 2.44          & 0.06           & -0.86        & -1.08          & 0.37          & 0.11                              \\
                                                                             & IMA                                     & -0.18         & 3.52          & 2.98           & 0.74         & 1.40           & 2.61          & 1.84                              \\
                                                                             & IHA + IMA                               & 1.48          & 3.66          & 3.07           & 0.56         & 1.28           & 2.67          & 2.12                              \\ 
      \\   
      \end{tabular}%
\\
\footnotesize{$\ast$ marks the cases where the task performance scores of certain proposed data points are lowered to have a monotonic rate-performance curve in order to make BD-rate calculation possible. BD-task is calculated with original values.}\\
\end{table*}

\section{Conclusions}
\label{sec:summary}
In this paper, we proposed a hybrid coding system NN-VVC, which combines the high performance of a machine-task-optimized learned image codec (LIC) and a state-of-the-art conventional video codec (CVC) conforming to the Versatile Video Coding (VVC) standard. It was shown that the important characteristics for machine task in the reconstructed images generated by the LIC could be transferred to the inter-frames when the LIC encoded intra-frames are used as reference frames in the CVC encoding. Furthermore, the Intra Human Adapter (IHA) is applied to the LIC encoded intra-frames to reduce the artefacts introduced by the LIC, resulting in a more efficient inter-frame coding, while keeping the machine-oriented characteristics. The decoded inter-frames are further adapted for machine consumption with a learned Intra Machine Adapter (IMA). The NN-VVC showed significant coding gains over the VVC codec in terms of machine task performance on similar bitrates.  Future research will focus on optimizing the NN-VVC system for both machine and human consumption.

\small
\bibliographystyle{template/IEEEtran/IEEEtran}
\bibliography{refs}

\end{document}